\documentclass[letter]{aa}     
\usepackage{indentfirst}
\usepackage{graphicx}
\usepackage{txfonts}
\usepackage{float}
\usepackage{lipsum}
\usepackage{subcaption}        
\usepackage{hyperref}
\hypersetup{
     colorlinks   = true,
     linkcolor   =  cyan,
     citecolor    = teal,
     urlcolor = blue
}
                               
\usepackage{lscape}             
\usepackage{xcolor}

\usepackage{placeins}           

\DeclareRobustCommand{\Mpc}{\mathrm{Mpc}}
\DeclareRobustCommand{\msun}{\mathrm{M}_{\odot}}

\DeclareRobustCommand{\kpc}{\mathrm{kpc}}

\DeclareRobustCommand{\magarcsec}{\mathrm{mag \ arcsec^{-2}}}

\begin{document}

   \title{Serendipitous discovery of an almost-dark galaxy \\ in the Virgo Cluster}

   \author{Minh Ngoc Le \inst{1,2,3}\fnmsep\thanks{Corresponding author: lmngoc1509@gmail.com}
        \and Ignacio Trujillo \inst{1,2}
        \and Sergio Guerra Arencibia \inst{1, 2}
        \and Ignacio Ruiz Cejudo \inst{1,2}
        \and Junais \inst{4}
        \and Mireia Montes \inst{7}
        \and Johan H. Knapen \inst{1,2}
        \and Željko Ivezić \inst{5}
        \and Michael H.F. Wilkinson \inst{6}
        \and Reynier Peletier \inst{3}
        \and Miquel Serra-Ricart \inst{8,1,2}
        \and Miguel R. Alarcon \inst{8,1,2}
        }

   \institute{Instituto de Astrofísica de Canarias, Vía Láctea S/N, E-38205 La Laguna, Spain 
   \and Departamento de Astrofísica, Universidad de La Laguna, E-38206 La Laguna, Spain
   \and Kapteyn Astronomical Institute, University of Groningen, P.O. Box 800, 9700AV Groningen, The Netherlands
   \and Leibniz-Institut für Astrophysik Potsdam (AIP), An der Sternwarte 16, 14482 Potsdam, Germany
   \and Department of Astronomy and the DiRAC Institute, University of Washington, 3910 15th Avenue, NE, Seattle, WA 98195, USA
   \and Bernoulli Institute of Mathematics, Computer Science and Artificial Intelligence, University of Groningen, Groningen, The Netherlands
   \and Institute of Space Sciences (ICE, CSIC), Campus UAB, Carrer de Can Magrans, s/n, E-08193 Barcelona, Spain
   \and Light Bridges, Observatorio del Teide, Carretera del Observatorio s/n, E-38500 Guimar, Tenerife, Canarias, Spain
   }
   
   \authorrunning{Le et al.}
   \titlerunning{An almost-dark galaxy in the Virgo Cluster}
   
   \date{\today}

  \abstract
   {Analogues of extreme Local Group galaxies with very low surface brightnesses and large effective radii must exist elsewhere, still hidden due to current detection limits. We report the serendipitous discovery of the low-mass almost-dark galaxy TTT~J1237327+143535 in the Virgo Cluster, observed using the Two-meter Twin Telescope at the Teide Observatory. TTT~J1237327+143535 has a central surface brightness of $27.9 \pm 0.2 \, \magarcsec{}$ in the $g'$ band, an effective radius of $0.9 \pm 0.1  \, \kpc{}$, and a total stellar mass of $(2.2 \pm 0.4) \times 10^6 \, \msun{}$. Its effective radius and absolute magnitude are similar to those of galaxies And~XXI and And~XXIII. The discovery of this extremely low-surface-brightness, extended, low-mass galaxy suggests the existence of a significant population of almost-dark galaxies in the Virgo Cluster. An in-depth analysis of the Next Generation Virgo Cluster Survey, as well as upcoming Rubin data releases and the 10-year Rubin Legacy Survey of Space and Time, is expected to reveal large samples of this extreme galaxy population, which will offer insights into galaxy formation in extreme conditions.}


   \maketitle
   \nolinenumbers

\section{Introduction}

There is still no consensus within the scientific community on what `almost-dark' galaxies are. One working definition is galaxies not detected by traditional sky surveys, such as the Sloan Digital Sky Survey (SDSS; \citealt{2000AJ....120.1579Y}). In practical terms, an almost-dark galaxy is one whose central surface brightness in the $r$ band is fainter than $26 \, \magarcsec{}$. Well-known examples of this type of galaxy include Nube \citep{Nube_2024A&A...681A..15M} and RCP~32 \citep{2021A&A...656A..44R}.

Almost-dark galaxies found in deeper surveys than the SDSS, such as Stripe 82 (e.g. \citealt{2016MNRAS.456.1359F}) and Dark Energy Camera Legacy Survey \citep{2019AJ....157..168D}, have effective radii similar to those of ultra-diffuse galaxies (UDGs), i.e. $r_\mathrm{e} > 1 \, \kpc{}$ \citep{2015ApJ...798L..45V}, but are significantly fainter, by a factor of more than $10$. Current galaxy simulations cannot reproduce galaxies with this combination of characteristics (see e.g. \citealt{2022NatAs...6..897S} and \citealt{2025MNRAS.541.1195R}). This raises many questions about the formation, abundance, and environments of these galaxies.

Star-counting techniques have enabled the detection of very low-surface-brightness galaxies in the Local Group, but their effective radii are in general significantly smaller ($r_\mathrm{e} < 1 \, \kpc$; \citealt{McConnachie_2012}). Galaxies of this type should certainly also exist outside the Local Group. However, their small apparent extensions and low surface brightnesses would make them indistinguishable from the background of high-redshift galaxies. Nonetheless, there are some galaxies in the Local Group with low surface brightnesses whose effective radius is around $1 \, \kpc{}$, such as And~XXI and And~XXIII. In this letter, we report the discovery of one of these almost-dark galaxies (TTT~J1237327+143535) in the Virgo Cluster. While it is not the first galaxy with a very low surface brightness found in this cluster (others have been reported in \citealt{Mihos_2015, Mihos_2017}, \citealt{2020ApJ...890..128F} and \citealt{2020ApJ...899...69L}), TTT~J1237327+143535 is one of the most extreme cases to date.

\section{Discovery imaging}\label{section:data}

The almost-dark galaxy TTT~J1237327+143535 was discovered during the construction of a control field for the galaxy Malin~1 with TTT3 (the first 2 m telescope of the TTT facility), a Ritchey-Chrétien 2 m telescope located at Teide Observatory, using  $g^\prime$-, $r^\prime$-, and $i^\prime$-band standard Sloan filters manufactured by Baader Planetarium. For the $g^\prime$- and  $r^\prime$-band observations, the instrument used was FERVOR-M with the sensor BSI sCMOS IMX455 \citep{2023PASP..135e5001A}, with a field of view of $10.2^{\prime}$ x $6.8^{\prime}$ and a pixel scale of $0.194^{\prime\prime}$ $\rm{pix}^{-1}$ ($3 \times 3$ binning). For the $i^\prime$ band, the instrument used was COLORS with the sensor BSI e2v CCD42-40 BEX2-DD, a field of view of $7.85^{\prime}$ x $7.85^{\prime}$ and a pixel scale of $0.23^{\prime\prime}$.

The observations were conducted between April and June of 2025. The exposure times for the $g^\prime$, $r^\prime$, and $i^\prime$ bands were $6.5$, $7.8$, and $5$ hours, respectively. The full width at half maximum for point-like sources on the stacked images is $1.3''$, $1.1''$, and $1.5''$, and the typical sky brightness during the observations was $21.7$, $20.85$, and $20.3 \, \magarcsec{}$ for the $g^\prime$, $r^\prime$, and $i^\prime$ bands, respectively. The surface brightness limits achieved in the final images are $30.3$, $29.6$, and $29.2 \, \magarcsec{}$ (in areas equivalent to $10^{\prime\prime}\times10^{\prime\prime}$ at a $3\sigma$ level) for the respective bands. The $5\sigma$ point source limits are at $26.2, 25.4$, and $24.8 \, \mathrm{mag}$ for $g^\prime$, $r^\prime$, and $i^\prime$ bands in apertures with radii of $1.5''$. To preserve the faint signal present in the data, we used a low-surface-brightness-compliant pipeline. A thorough explanation of the observational and reduction strategy is in \cite{junais2025}.

\begin{figure*}[ht!]
    \sidecaption
    {\includegraphics[width=12cm] {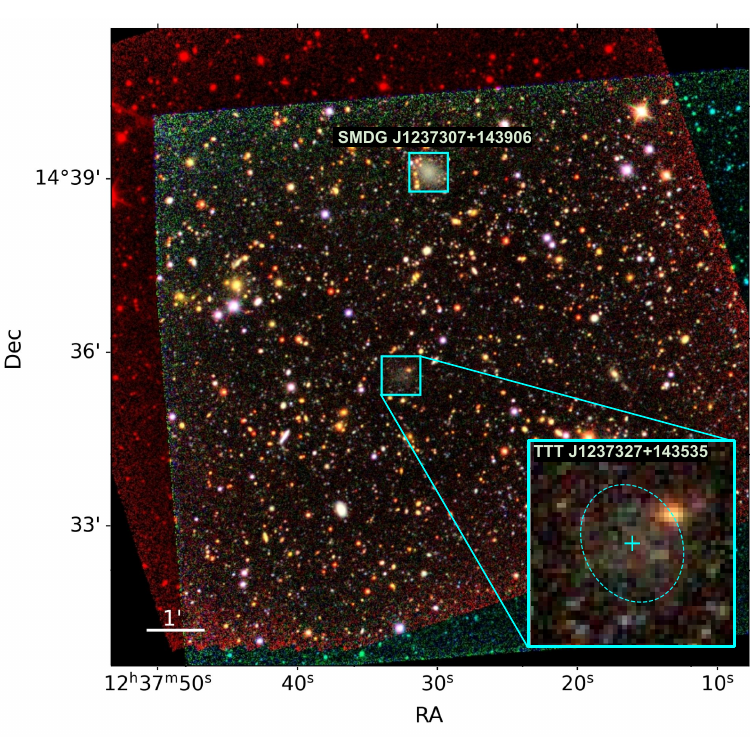}}
      \caption{Colour image (combined $g'$, $r'$, and $i'$ filters, using \textit{astscript-color-faint-gray} \citep{2024RNAAS...8...10I}) of the almost-dark galaxy TTT~J1237327+143535 (outlined by the lower blue square; the cross marks the centre of the object and the dashed ellipse shows the half-light radius) and its neighbour SMDG~J1237307+143906 (upper blue square). The boxes around these objects mark areas of $40''\times40''$ ($3 \, \kpc{} \times 3 \, \kpc{}$ at the Virgo Cluster distance)}
    \label{fig:dwarfs color image}
    \end{figure*}

Figure~\ref{fig:dwarfs color image} shows a colour image that combines the $g'$-, $r'$-, and $i'$-band images of the almost-dark galaxy TTT~J1237327+143535, together with the UDG neighbour SMDG~J1237307+143906 reported in the Systematically Measuring Ultra-diffuse Galaxies (SMUDGes) survey \citep{2023ApJS..267...27Z}. Giving their positions in the sky (at a projected distance of $0.77 \, \Mpc{}$ from M87) and their extensions, we assumed that both the almost-dark galaxy and the UDG neighbour are associated with the Virgo Cluster, at $\mathrm{D} = 16.17 \, \Mpc{}$, with a statistical error of $ \pm 0.25 \, \Mpc{}$  and a systematical error of $\pm 0.47 \, \ \Mpc{} $ \citep{2025ApJ...982...26A}.

\section{Physical properties}\label{section:analysis}

We derived the physical properties of TTT~J1237327+143535 and SMDG~J1237307+143906 by first running the source detection algorithm MTO2 \citep{MTO2} to detect and mask all the other sources in a region of $5'\times 5'$ around the galaxy. We then used Photutils \citep{bradley_2026_19636730} to find the galaxy centre by fitting isophotes on the masked image, and GALFIT \citep{galfit2010AJ....139.2097P} to fit the galaxy with a Sérsic model and derive the Sérsic index, effective radius, ellipticity, and position angle.  

TTT~J1237327+143535 has a rather flat radial surface brightness profile with Sérsic index $n = 0.54 \pm 0.03$, while the neighbour has a slightly steeper profile with $n = 0.75 \pm 0.01$. The two galaxies have similar effective radii of around $12''$, which corresponds to $1 \, \kpc{}$ at the distance of the Virgo Cluster. The other properties of these objects are listed in Table~\ref{tab:dwarfs properties}. 

  We followed \cite{1998ApJ...500..525S} and \cite{2011ApJ...737..103S} and applied an extinction correction for each band. TTT~J1237327+143535 (Fig~\ref{fig:profiles}, top panel) has an extremely faint central surface brightness of $27.9 \, \magarcsec{} $ and $26.9 \, \magarcsec{} $ in the $g'$ (blue squares) and $r'$ bands (orange squares), respectively. SMDG~J1237307+143906 (same panel, circles), in comparison, has a central surface brightness of $25.4 \, \magarcsec{}$ and  $24.8 \, \magarcsec{}$ in the $g'$ and $r'$ bands, two magnitudes brighter than its almost-dark neighbour. TTT~J1237327+143535 has a median extinction-corrected colour of $(g'-r')_0 = 0.6 \pm 0.1$ within $r_\mathrm{e}$. The UDG neighbour is also similarly red, with $(g'-r')_0 = 0.58 \pm 0.04$. 
 
 From these colours, we derived a colour-mass-to-light ratio for the $g'$ band of $\log_{10} (M_\star/L) = m \times (g'-r')_0 + b,$ assuming a Kroupa initial mass function \citep{2001MNRAS.322..231K} and the values of $m = 2.029$ and $b = -0.984$ from \cite{Roediger_15_M_L}. Following \cite{Bakos2008ApJ...683L.103B}, the stellar mass density profile was obtained as $\log_{10} \Sigma_\star(r) = \log_{10}({M_\star/L}) - 0.4[\mu_{g'}(r) - m_{g',\odot}] + 8.629,$ where the absolute magnitude of the Sun is $m_{g',\odot} = 5.11 \, \mathrm{mag}$. To avoid the effect of the noisy outskirts of TTT~J1237327+143535, we integrated the surface mass density up to its effective radius, then multiplied it by 2 to obtain the total stellar mass. 

 We find that both galaxies have very low surface mass densities. SMDG~J1237307+143906 (Fig~\ref{fig:profiles}, bottom panel, circles) has a central mass surface density of $5 \, \mathrm{M_\odot \, pc^{-2}} $ which drops down to almost $0.1 \, \mathrm{M_\odot \, pc^{-2}}$ at around two times its effective radius ($\sim 2\, \kpc{}$). Its total stellar mass is about $10^7 \, \msun{}$. In the almost-dark galaxy (squares), even at its centre the surface mass density is below $1 \, \mathrm{M_\odot \, pc^{-2}} $: it is $0.8 \, \mathrm{M_\odot \, pc^{-2}} $, and drops to $0.1 \, \mathrm{M_\odot \, pc^{-2}} $ at the outskirts. It has a total stellar mass of $2 \times 10^6 \, \msun{}$, which is five times less than its UDG neighbour.

\section{Discussion}\label{section:discussions}

\begin{figure*}[ht!]
    \centering
    {\includegraphics[width=1\linewidth] {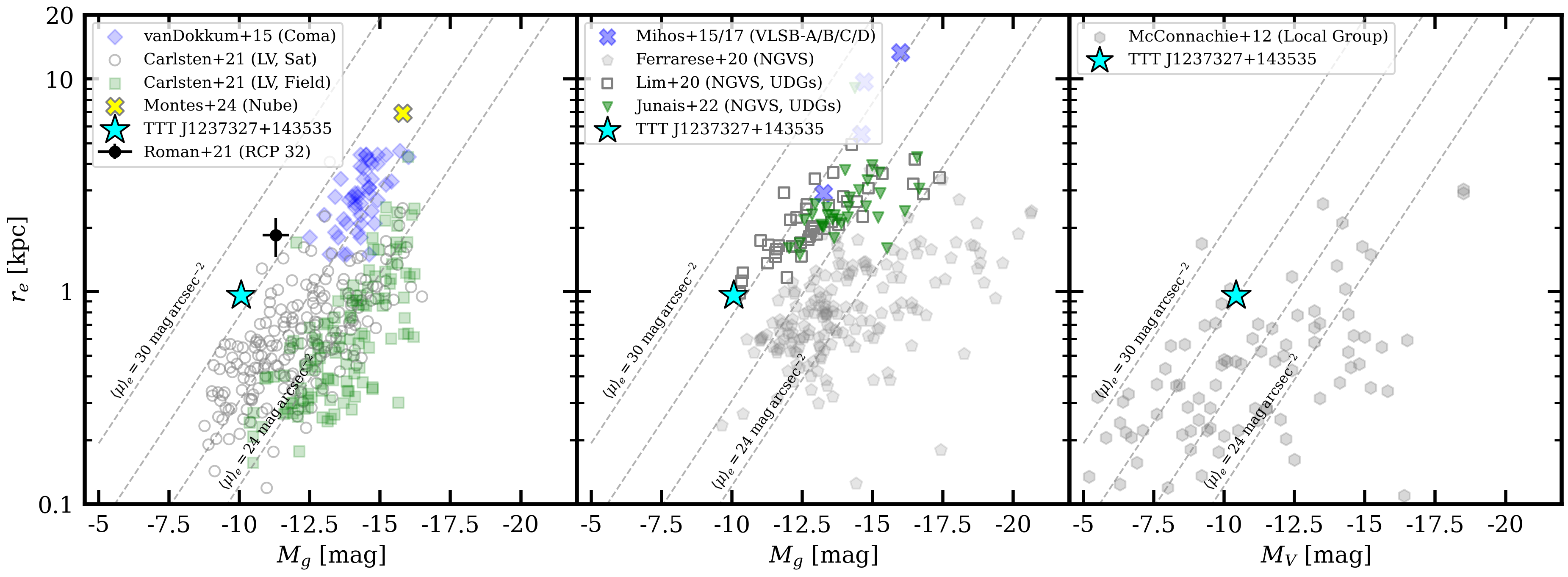}}
      \caption{\textit{Left}: Absolute magnitude in the $g$ band and effective radius of the almost-dark galaxy TTT~J1237327+143535 (marked with a cyan star), compared to UDGs in the Coma Cluster (diamonds; \citealt{2015ApJ...798L..45V}), satellite (open circles) and field (squares) dwarfs in the Local Volume (D $\leq 12 \, \Mpc{}$; \citealt{Carlsten2021ApJ...922..267C}), and two other almost-dark galaxies Nube (cross) and RCP~32 (black circle with an errorbar). \textit{Middle}: Same but compared to other objects in the Virgo Cluster from the NGVS: galaxies with \texttt{certain} membership within $0.3 \, \Mpc{}$ of M87 (\citealt{2020ApJ...890..128F}, pentagons), and UDGs from \citealt{2022A&A...667A..76J} (triangles), and in \citealt{2020ApJ...899...69L} (squares). We also show the low-surface-brightness galaxies found with the Burrell Schmidt Deep Virgo Survey for comparison (\citealt{Mihos_2015, Mihos_2017}, blue crosses). \textit{Right}: Absolute magnitude in the $V$ band versus effective radius of TTT~J1237327+143535 (cyan star) compared to galaxies in the Local Group (D $\leq 3 \, \Mpc{}$, \citealt{McConnachie_2012}; hexagons). The dashed grey lines show the corresponding effective surface brightness at each magnitude -- effective radius. Error bars of the measurement for TTT~J1237327+143535 are small and not visible in the plot.}
    \label{fig:compared_to_literature}
\end{figure*}

      The distance to TTT~J1237327+143535 is unknown. There are two reasons to support the assumption that it is associated with the Virgo Cluster. Firstly, the sky separation between TTT~J1237327+143535 and the giant elliptical galaxy M87 at the centre of the Virgo Cluster is $2.74^\circ$, corresponding to $0.77 \, \Mpc{}$ at the Virgo distance. Thus, it is located within the virial radius of the Virgo Cluster ($r_{\mathrm{vir}} = 1.55 \, \Mpc{}$, from \citealt{2012ApJS..200....4F}). Secondly, the fact that we do not resolve the stellar populations of TTT~J1237327+143535 is proof of its position beyond the Local Group (within $3 \, \Mpc{}$; \citealt{McConnachie_2012}). On the other hand, it is unlikely that TTT~J1237327+143535 is farther away than the Virgo Cluster. In fact, if it were at about $30 \, \Mpc{}$, its effective radius could be twice the current radius, and its total stellar mass would be $3.5$ times higher than the values measured at the Virgo distance. Such an object with a central surface brightness of $28 \, \magarcsec{}$ would be even harder to explain with current simulations of galaxy formation and evolution.

    We find no signal from the HI spectrum at the position of TTT~J1237327+143535 from the ALFALFA survey. Based on this, we place an upper limit on its HI mass of $10^7 \, \msun{}$ at the Virgo Cluster distance \citep{2007AJ....133.2569G}. Furthermore, there is also no H$\alpha$ reported at the same location when cross-matched with the Virgo Environmental Survey Tracing Ionised Gas Emission survey (VESTIGE; \citealt{2018A&A...614A..56B}). A galaxy with such a faint surface brightness and relatively red colour may indeed be HI-poor and have no recent star-forming activity.  

   TTT~J1237327+143535 is a remarkable object with an extremely low density of baryonic matter, as reflected in its low surface brightness and low stellar surface mass density. It is clearly distinguished as the most extended galaxy among those dwarfs and UDGs with similar absolute magnitudes in the Local Volume (Fig~\ref{fig:compared_to_literature}, left panel). It is about twice as large in effective radius as the other dwarfs with \texttt{certain} membership in the Virgo Cluster (middle panel). TTT~J1237327+143535 and two other almost-dark galaxies, Nube and RCP~32, all have large effective radii compared to galaxies of the same magnitudes. Additionally, its effective radius is comparable to those of the UDGs (as defined in \citealt{2015ApJ...798L..45V}), but it is $100$ times fainter than UDGs. Current state-of-the-art simulations (e.g \citealt{2025MNRAS.541.1195R}) can reproduce galaxies at the same magnitude as TTT~J1237327+143535 with half or even one-third of its effective radius. Thus, a larger sample of similarly faint and extended galaxies will provide an observational constraint on the physical mechanisms responsible for the extensive radii of this population.
    
  TTT~J1237327+143535 occupies a position on the effective radius--magnitude space that remains genuinely rare even in the deepest existing surveys of the Virgo Cluster (Fig.\ref{fig:compared_to_literature}, middle panel). The UDG catalog of the NGVS \citep{2020ApJ...899...69L} -- one of the most comprehensive low-surface-brightness surveys of Virgo to date, covering 104 degrees squared -- reveals only three objects (NGVSUDG-22, NGVSUDG-A06, and NGVSUDG-A12) with comparable properties, all at the extreme faint end of their sample; all three are those close to the NGVS detection and completeness limit ($50 \%$ complete at $M_g = -9.6$). TTT~J1237327+143535, discovered independently, naturally complements and reinforces the NGVS~UDG sample by confirming that objects in this extreme parameter space exist. Finding a galaxy with those extreme properties by chance, in an observation independent of the NGVS, reinforces the idea that there is a population of extended, low-mass, low-surface-brightness objects in the Virgo Cluster. It is worth emphasizing how difficult it is to find an object with these characteristics. Although the NGVS deep imaging covers the location of TTT~J1237327+143535, the galaxy is not included in the NGVS catalog.

     We consider it unlikely that there is a globular cluster (GC) associated with  TTT~J1237327+143535. The point-source limit of our data corresponds to $M_g \sim -5 \, \mathrm{mag}$ at the Virgo distance, which is $2 \, \mathrm{mag}$ fainter than the peak of GC luminosity function ( $M_g \sim -7.2 \, \mathrm{mag}$, as in \citealt{2020ApJ...899...69L}). However, due to the limitations of ground-based imaging in separating GCs from contaminants, and to avoid the assumption of a universal GC luminosity function, especially for low-mass, low surface brightness galaxies such as TTT~J1237327+143535, space-based follow-up with $u$-band coverage (the most effective wavelength range for discriminating GC candidates from stars; \citealt{2014ApJS..210....4M}) would enable a robust characterisation of the GC system of this almost-dark galaxy.
    
    TTT~J1237327+143535 has a similar absolute magnitude and effective radius as the dwarf galaxies at the faint and low-mass end of the Local Group sample from \citealt{McConnachie_2012} (Fig~\ref{fig:compared_to_literature}, right panel)\footnote{$V$-band magnitude is converted from $g'$ and $r'$ bands via $V = g'-0.5784(g'-r')-0.0038$, following \url{https://classic.sdss.org/dr4/algorithms/sdssUBVRITransform.php##Lupton2005}.}. The closest Local Group analogues to our almost-dark galaxy are the And~XXI and And~XXIII satellites of Andromeda. Finding an analogue to these satellites in the Virgo Cluster by chance suggests the existence of a significant population of extremely faint and low-mass galaxies beyond the Local Group. This population remains undetected due to the current survey surface brightness limits. A dedicated search in optical surveys -- such as the Large Binocular Telescope Imaging of Galactic Halos and Tidal Structures survey (\citealt{Trujillo_LIGHTS_2021} and \citealt{2024AJ....168...69Z}), the NGVS \citep{2012ApJS..200....4F}, the Vera Rubin Legacy Survey of Space and Time (LSST) \citep{2019ApJ...873..111I}, and those carried out by Euclid \citep{2025A&A...697A...1E} and the Roman Space Telescope \citep{2023arXiv230609414M} -- will certainly reveal more about this very faint population.

\section{Conclusions}\label{section:conclusions}

In this letter, we have reported the discovery of the almost-dark galaxy TTT~J1237327+143535 in the Virgo Cluster. This galaxy has an extremely low central surface brightness -- of $28 \, \magarcsec{}$ in the $g'$ band, an effective radius of $1 \, \kpc{}$, and a total stellar mass of $2 \times 10^6 \, \msun{}$. These properties make TTT~J1237327+143535 remarkable in comparison to dwarfs and UDGs beyond the Local Group and in the Virgo Cluster. The galaxy is similar to the two satellite galaxies And~XXI and And~XXIII in the Local Group.

The serendipitous discovery of TTT~J1237327+143535 suggests that there are more objects with similar properties at distances beyond the Local Group. Upcoming surveys with wide-area coverage and similar depth will potentially detect many more almost-dark galaxies. Finding more galaxies of this type can help determine how galaxies form and evolve under extreme physical conditions.

\begin{acknowledgements}
      Co-funded by the European Union (MSCA Doctoral Network EDUCADO, GA 101119830, Widening Participation, ExGal-Twin, GA 101158446, and UNDARK, GA 101159929). MNL thanks Betsey Adams, Javier Roman, and Dennis Zaritsky for helpful discussions. IT and ICR acknowledge grant PID2022-140869NB-I00 and IAC project P/302302, financed by MCIN. J acknowledges the Alexander von Humboldt Foundation through a Humboldt Research Fellowship. MM acknowledges grants RYC2022-036949-I financed by the MICIU/AEI/10.13039/501100011033 and by ESF+, and program Unidad de Excelencia Mar\'{i}a de Maeztu CEX2020-001058-M, financed by MCIN/AEI/10.13039/501100011033, and by the MaX-CSIC Excellence Award MaX4-SOMMA-ICE. JHK acknowledges grants PID2022-136505NB-I00, funded by MCIN/AEI/10.13039/501100011033 and EU, ERDF. This work is part of grant CEX2025-001609-S, awarded to the IAC under the Severo Ochoa Centre of Excellence program and funded by MICIU/AEI/10.13039/501100011033. The research utilized Observing Time Rights and Indefeasible Computer Rights in the PEI "GALAXDIF25", with storage and computing capacity at ASTRO POC’s EDGE, in collaboration with Hewlett Packard Enterprise and VAST DATA.
\end{acknowledgements}

\bibliographystyle{bibtex/aa}
\bibliography{references.bib}

@ARTICLE{Trujillo_LIGHTS_2021,
       author = {{Trujillo}, Ignacio and {D'Onofrio}, Mauro and {Zaritsky}, Dennis and {Madrigal-Aguado}, Alberto and {Chamba}, Nushkia and {Golini}, Giulia and {Akhlaghi}, Mohammad and {Sharbaf}, Zahra and {Infante-Sainz}, Ra{\'u}l and {Rom{\'a}n}, Javier and {Morales-Socorro}, Carlos and {Sand}, David J. and {Martin}, Garreth},
        title = "{Introducing the LBT Imaging of Galactic Halos and Tidal Structures (LIGHTS) survey. A preview of the low surface brightness Universe to be unveiled by LSST}",
      journal = {\aap},
         year = 2021,
        month = oct,
       volume = {654},
          eid = {A40},
        pages = {A40},
          doi = {10.1051/0004-6361/202141603},
archivePrefix = {arXiv},
       eprint = {2109.07478},
 primaryClass = {astro-ph.GA},
       adsurl = {https://ui.adsabs.harvard.edu/abs/2021A&A...654A..40T}
}

@ARTICLE{2019ApJ...873..111I,
       author = {{Ivezi{\'c}}, {\v{Z}}eljko and {Kahn}, Steven M. and {Tyson}, J. Anthony and {Abel}, Bob and {Acosta}, Emily and {Allsman}, Robyn and {Alonso}, David and {AlSayyad}, Yusra and {Anderson}, Scott F. and {Andrew}, John and {Angel}, James Roger P. and {Angeli}, George Z. and {Ansari}, Reza and {Antilogus}, Pierre and {Araujo}, Constanza and {Armstrong}, Robert and {Arndt}, Kirk T. and {Astier}, Pierre and {Aubourg}, {\'E}ric and {Auza}, Nicole and {Axelrod}, Tim S. and {Bard}, Deborah J. and {Barr}, Jeff D. and {Barrau}, Aurelian and {Bartlett}, James G. and {Bauer}, Amanda E. and {Bauman}, Brian J. and {Baumont}, Sylvain and {Bechtol}, Ellen and {Bechtol}, Keith and {Becker}, Andrew C. and {Becla}, Jacek and {Beldica}, Cristina and {Bellavia}, Steve and {Bianco}, Federica B. and {Biswas}, Rahul and {Blanc}, Guillaume and {Blazek}, Jonathan and {Blandford}, Roger D. and {Bloom}, Josh S. and {Bogart}, Joanne and {Bond}, Tim W. and {Booth}, Michael T. and {Borgland}, Anders W. and {Borne}, Kirk and {Bosch}, James F. and {Boutigny}, Dominique and {Brackett}, Craig A. and {Bradshaw}, Andrew and {Brandt}, William Nielsen and {Brown}, Michael E. and {Bullock}, James S. and {Burchat}, Patricia and {Burke}, David L. and {Cagnoli}, Gianpietro and {Calabrese}, Daniel and {Callahan}, Shawn and {Callen}, Alice L. and {Carlin}, Jeffrey L. and {Carlson}, Erin L. and {Chandrasekharan}, Srinivasan and {Charles-Emerson}, Glenaver and {Chesley}, Steve and {Cheu}, Elliott C. and {Chiang}, Hsin-Fang and {Chiang}, James and {Chirino}, Carol and {Chow}, Derek and {Ciardi}, David R. and {Claver}, Charles F. and {Cohen-Tanugi}, Johann and {Cockrum}, Joseph J. and {Coles}, Rebecca and {Connolly}, Andrew J. and {Cook}, Kem H. and {Cooray}, Asantha and {Covey}, Kevin R. and {Cribbs}, Chris and {Cui}, Wei and {Cutri}, Roc and {Daly}, Philip N. and {Daniel}, Scott F. and {Daruich}, Felipe and {Daubard}, Guillaume and {Daues}, Greg and {Dawson}, William and {Delgado}, Francisco and {Dellapenna}, Alfred and {de Peyster}, Robert and {de Val-Borro}, Miguel and {Digel}, Seth W. and {Doherty}, Peter and {Dubois}, Richard and {Dubois-Felsmann}, Gregory P. and {Durech}, Josef and {Economou}, Frossie and {Eifler}, Tim and {Eracleous}, Michael and {Emmons}, Benjamin L. and {Fausti Neto}, Angelo and {Ferguson}, Henry and {Figueroa}, Enrique and {Fisher-Levine}, Merlin and {Focke}, Warren and {Foss}, Michael D. and {Frank}, James and {Freemon}, Michael D. and {Gangler}, Emmanuel and {Gawiser}, Eric and {Geary}, John C. and {Gee}, Perry and {Geha}, Marla and {Gessner}, Charles J.~B. and {Gibson}, Robert R. and {Gilmore}, D. Kirk and {Glanzman}, Thomas and {Glick}, William and {Goldina}, Tatiana and {Goldstein}, Daniel A. and {Goodenow}, Iain and {Graham}, Melissa L. and {Gressler}, William J. and {Gris}, Philippe and {Guy}, Leanne P. and {Guyonnet}, Augustin and {Haller}, Gunther and {Harris}, Ron and {Hascall}, Patrick A. and {Haupt}, Justine and {Hernandez}, Fabio and {Herrmann}, Sven and {Hileman}, Edward and {Hoblitt}, Joshua and {Hodgson}, John A. and {Hogan}, Craig and {Howard}, James D. and {Huang}, Dajun and {Huffer}, Michael E. and {Ingraham}, Patrick and {Innes}, Walter R. and {Jacoby}, Suzanne H. and {Jain}, Bhuvnesh and {Jammes}, Fabrice and {Jee}, M. James and {Jenness}, Tim and {Jernigan}, Garrett and {Jevremovi{\'c}}, Darko and {Johns}, Kenneth and {Johnson}, Anthony S. and {Johnson}, Margaret W.~G. and {Jones}, R. Lynne and {Juramy-Gilles}, Claire and {Juri{\'c}}, Mario and {Kalirai}, Jason S. and {Kallivayalil}, Nitya J. and {Kalmbach}, Bryce and {Kantor}, Jeffrey P. and {Karst}, Pierre and {Kasliwal}, Mansi M. and {Kelly}, Heather and {Kessler}, Richard and {Kinnison}, Veronica and {Kirkby}, David and {Knox}, Lloyd and {Kotov}, Ivan V. and {Krabbendam}, Victor L. and {Krughoff}, K. Simon and {Kub{\'a}nek}, Petr and {Kuczewski}, John and {Kulkarni}, Shri and {Ku}, John and {Kurita}, Nadine R. and {Lage}, Craig S. and {Lambert}, Ron and {Lange}, Travis and {Langton}, J. Brian and {Le Guillou}, Laurent and {Levine}, Deborah and {Liang}, Ming and {Lim}, Kian-Tat and {Lintott}, Chris J. and {Long}, Kevin E. and {Lopez}, Margaux and {Lotz}, Paul J. and {Lupton}, Robert H. and {Lust}, Nate B. and {MacArthur}, Lauren A. and {Mahabal}, Ashish and {Mandelbaum}, Rachel and {Markiewicz}, Thomas W. and {Marsh}, Darren S. and {Marshall}, Philip J. and {Marshall}, Stuart and {May}, Morgan and {McKercher}, Robert and {McQueen}, Michelle and {Meyers}, Joshua and {Migliore}, Myriam and {Miller}, Michelle and {Mills}, David J.},
        title = "{LSST: From Science Drivers to Reference Design and Anticipated Data Products}",
      journal = {\apj},
         year = 2019,
        month = mar,
       volume = {873},
       number = {2},
          eid = {111},
        pages = {111},
          doi = {10.3847/1538-4357/ab042c},
archivePrefix = {arXiv},
       eprint = {0805.2366},
 primaryClass = {astro-ph},
       adsurl = {https://ui.adsabs.harvard.edu/abs/2019ApJ...873..111I}
}

@ARTICLE{junais2025,
       author = {{Junais} and {Ruiz Cejudo}, Ignacio and {Guerra Arencibia}, Sergio and {Trujillo}, Ignacio and {Alarcon}, Miguel R. and {Serra-Ricart}, Miquel and {Knapen}, Johan H. and {Duc}, Pierre-Alain},
        title = "{Deep imaging of the galaxy Malin 2 shows new faint structures and a candidate satellite dwarf galaxy}",
      journal = {\aap},
         year = 2025,
        month = oct,
       volume = {702},
          eid = {A136},
        pages = {A136},
          doi = {10.1051/0004-6361/202556569},
archivePrefix = {arXiv},
       eprint = {2508.07930},
 primaryClass = {astro-ph.GA},
       adsurl = {https://ui.adsabs.harvard.edu/abs/2025A&A...702A.136J}
}

@ARTICLE{MTO2,
  author={Hashem Faezi, Mohammad and Peletier, Reynier and Wilkinson, Michael H. F.},
  journal={IEEE Access}, 
  title={Multi-Spectral Source-Segmentation Using Semantically-Informed Max-Trees}, 
  year={2024},
  volume={12},
  number={},
  pages={72288-72302},
  doi={10.1109/ACCESS.2024.3403309}
}

@software{bradley_2026_19636730,
  author       = {Bradley, Larry and
                  Sipőcz, Brigitta M. and
                  Robitaille, T. P. and
                  Tollerud, E. J. and
                  Vinícius, Zé and
                  Deil, Christoph and
                  Barbary, Kyle and
                  Wilson, Tom J. and
                  Busko, Ivo and
                  Donath, Axel and
                  Günther, Hans Moritz and
                  Cara, Mihai and
                  Lim, P. L. and
                  Meßlinger, Sebastian and
                  Conseil, Simon and
                  Droettboom, Michael and
                  Bostroem, K. Azalee and
                  Bray, E. M. and
                  Bratholm, Lars Andersen and
                  Burnett, Zach and
                  Jamieson, William and
                  Ginsburg, Adam and
                  Taranu, Dan and
                  Barentsen, Geert and
                  Craig, Matthew W. and
                  Morris, Brett M. and
                  Perrin, Marshall and
                  Rathi, Shivangee},
  title        = {Photutils},
  month        = apr,
  year         = 2026,
  publisher    = {Zenodo},
  version      = {3.0.0},
  doi          = {10.5281/zenodo.19636730},
  url          = {https://doi.org/10.5281/zenodo.19636730},
  swhid        = {swh:1:dir:5ea90432cd987336a26af201c1b50343b12e2daa
                   ;origin=https://doi.org/10.5281/zenodo.596036;visi
                   t=swh:1:snp:25dd84d95353ec3813baa8785c248017da7192
                   33;anchor=swh:1:rel:253cbde27398f9aea9d7368f386661
                   8c00fe16f9;path=astropy-photutils-3322558
                  },
}

@ARTICLE{2011ApJ...737..103S,
       author = {{Schlafly}, Edward F. and {Finkbeiner}, Douglas P.},
        title = "{Measuring Reddening with Sloan Digital Sky Survey Stellar Spectra and Recalibrating SFD}",
      journal = {\apj},
         year = 2011,
        month = aug,
       volume = {737},
       number = {2},
          eid = {103},
        pages = {103},
          doi = {10.1088/0004-637X/737/2/103},
archivePrefix = {arXiv},
       eprint = {1012.4804},
 primaryClass = {astro-ph.GA},
       adsurl = {https://ui.adsabs.harvard.edu/abs/2011ApJ...737..103S}
}

@ARTICLE{1998ApJ...500..525S,
       author = {{Schlegel}, David J. and {Finkbeiner}, Douglas P. and {Davis}, Marc},
        title = "{Maps of Dust Infrared Emission for Use in Estimation of Reddening and Cosmic Microwave Background Radiation Foregrounds}",
      journal = {\apj},
         year = 1998,
        month = jun,
       volume = {500},
       number = {2},
        pages = {525-553},
          doi = {10.1086/305772},
archivePrefix = {arXiv},
       eprint = {astro-ph/9710327},
 primaryClass = {astro-ph},
       adsurl = {https://ui.adsabs.harvard.edu/abs/1998ApJ...500..525S}
}

@ARTICLE{2001MNRAS.322..231K,
       author = {{Kroupa}, Pavel},
        title = "{On the variation of the initial mass function}",
      journal = {\mnras},
         year = 2001,
        month = apr,
       volume = {322},
       number = {2},
        pages = {231-246},
          doi = {10.1046/j.1365-8711.2001.04022.x},
archivePrefix = {arXiv},
       eprint = {astro-ph/0009005},
 primaryClass = {astro-ph},
       adsurl = {https://ui.adsabs.harvard.edu/abs/2001MNRAS.322..231K}
}

@article{Roediger_15_M_L,
    author = {Roediger, Joel C. and Courteau, Stéphane},
    title = {On the uncertainties of stellar mass estimates via colour measurements},
    journal = {Monthly Notices of the Royal Astronomical Society},
    volume = {452},
    number = {3},
    pages = {3209-3225},
    year = {2015},
    month = {07},
    issn = {0035-8711},
    doi = {10.1093/mnras/stv1499},
    url = {https://doi.org/10.1093/mnras/stv1499},
    eprint = {https://academic.oup.com/mnras/article-pdf/452/3/3209/4929279/stv1499.pdf},
}

@ARTICLE{Bakos2008ApJ...683L.103B,
       author = {{Bakos}, Judit and {Trujillo}, Ignacio and {Pohlen}, Michael},
        title = "{Color Profiles of Spiral Galaxies: Clues on Outer-Disk Formation Scenarios}",
      journal = {\apjl},
         year = 2008,
        month = aug,
       volume = {683},
       number = {2},
        pages = {L103},
          doi = {10.1086/591671},
archivePrefix = {arXiv},
       eprint = {0807.2776},
 primaryClass = {astro-ph},
       adsurl = {https://ui.adsabs.harvard.edu/abs/2008ApJ...683L.103B}
}

@ARTICLE{2025ApJ...982...26A,
       author = {{Anand}, Gagandeep S. and {Tully}, R. Brent and {Cohen}, Yotam and {Shaya}, Edward J. and {Makarov}, Dmitry I. and {Makarova}, Lidia N. and {Chazov}, Maksim I. and {Blakeslee}, John P. and {Cantiello}, Michele and {Jensen}, Joseph B. and {Kourkchi}, Ehsan and {Raimondo}, Gabriella},
        title = "{The TRGB─SBF Project. II. Resolving the Virgo Cluster with JWST}",
      journal = {\apj},
         year = 2025,
        month = mar,
       volume = {982},
       number = {1},
          eid = {26},
        pages = {26},
          doi = {10.3847/1538-4357/adb399},
archivePrefix = {arXiv},
       eprint = {2408.16810},
 primaryClass = {astro-ph.GA},
       adsurl = {https://ui.adsabs.harvard.edu/abs/2025ApJ...982...26A}
}

@ARTICLE{2023ApJS..267...27Z,
       author = {{Zaritsky}, Dennis and {Donnerstein}, Richard and {Dey}, Arjun and {Karunakaran}, Ananthan and {Kadowaki}, Jennifer and {Khim}, Donghyeon J. and {Spekkens}, Kristine and {Zhang}, Huanian},
        title = "{Systematically Measuring Ultra-diffuse Galaxies (SMUDGes). V. The Complete SMUDGes Catalog and the Nature of Ultradiffuse Galaxies}",
      journal = {\apjs},
         year = 2023,
        month = aug,
       volume = {267},
       number = {2},
          eid = {27},
        pages = {27},
          doi = {10.3847/1538-4365/acdd71},
archivePrefix = {arXiv},
       eprint = {2306.01524},
 primaryClass = {astro-ph.GA},
       adsurl = {https://ui.adsabs.harvard.edu/abs/2023ApJS..267...27Z}
}

@ARTICLE{galfit2010AJ....139.2097P,
       author = {{Peng}, Chien Y. and {Ho}, Luis C. and {Impey}, Chris D. and {Rix}, Hans-Walter},
        title = "{Detailed Decomposition of Galaxy Images. II. Beyond Axisymmetric Models}",
      journal = {\aj},
         year = 2010,
        month = jun,
       volume = {139},
       number = {6},
        pages = {2097-2129},
          doi = {10.1088/0004-6256/139/6/2097},
archivePrefix = {arXiv},
       eprint = {0912.0731},
 primaryClass = {astro-ph.CO},
       adsurl = {https://ui.adsabs.harvard.edu/abs/2010AJ....139.2097P}
}

@ARTICLE{Nube_2024A&A...681A..15M,
       author = {{Montes}, Mireia and {Trujillo}, Ignacio and {Karunakaran}, Ananthan and {Infante-Sainz}, Ra{\'u}l and {Spekkens}, Kristine and {Golini}, Giulia and {Beasley}, Michael and {Cebri{\'a}n}, Maria and {Chamba}, Nushkia and {D'Onofrio}, Mauro and {Kelvin}, Lee and {Rom{\'a}n}, Javier},
        title = "{An almost dark galaxy with the mass of the Small Magellanic Cloud}",
      journal = {\aap},
         year = 2024,
        month = jan,
       volume = {681},
          eid = {A15},
        pages = {A15},
          doi = {10.1051/0004-6361/202347667},
archivePrefix = {arXiv},
       eprint = {2310.12231},
 primaryClass = {astro-ph.GA},
       adsurl = {https://ui.adsabs.harvard.edu/abs/2024A&A...681A..15M}
}

@ARTICLE{2025A&A...697A...1E,
       author = {{Euclid Collaboration} and {Mellier}, Y. and {Abdurro'uf} and {Acevedo Barroso}, J.~A. and {Ach{\'u}carro}, A. and {Adamek}, J. and {Adam}, R. and {Addison}, G.~E. and {Aghanim}, N. and {Aguena}, M. and {Ajani}, V. and {Akrami}, Y. and {Al-Bahlawan}, A. and {Alavi}, A. and {Albuquerque}, I.~S. and {Alestas}, G. and {Alguero}, G. and {Allaoui}, A. and {Allen}, S.~W. and {Allevato}, V. and {Alonso-Tetilla}, A.~V. and {Altieri}, B. and {Alvarez-Candal}, A. and {Alvi}, S. and {Amara}, A. and {Amendola}, L. and {Amiaux}, J. and {Andika}, I.~T. and {Andreon}, S. and {Andrews}, A. and {Angora}, G. and {Angulo}, R.~E. and {Annibali}, F. and {Anselmi}, A. and {Anselmi}, S. and {Arcari}, S. and {Archidiacono}, M. and {Aric{\`o}}, G. and {Arnaud}, M. and {Arnouts}, S. and {Asgari}, M. and {Asorey}, J. and {Atayde}, L. and {Atek}, H. and {Atrio-Barandela}, F. and {Aubert}, M. and {Aubourg}, E. and {Auphan}, T. and {Auricchio}, N. and {Aussel}, B. and {Aussel}, H. and {Avelino}, P.~P. and {Avgoustidis}, A. and {Avila}, S. and {Awan}, S. and {Azzollini}, R. and {Baccigalupi}, C. and {Bachelet}, E. and {Bacon}, D. and {Baes}, M. and {Bagley}, M.~B. and {Bahr-Kalus}, B. and {Balaguera-Antolinez}, A. and {Balbinot}, E. and {Balcells}, M. and {Baldi}, M. and {Baldry}, I. and {Balestra}, A. and {Ballardini}, M. and {Ballester}, O. and {Balogh}, M. and {Ba{\~n}ados}, E. and {Barbier}, R. and {Bardelli}, S. and {Baron}, M. and {Barreiro}, T. and {Barrena}, R. and {Barriere}, J.-C. and {Barros}, B.~J. and {Barthelemy}, A. and {Bartolo}, N. and {Basset}, A. and {Battaglia}, P. and {Battisti}, A.~J. and {Baugh}, C.~M. and {Baumont}, L. and {Bazzanini}, L. and {Beaulieu}, J.-P. and {Beckmann}, V. and {Belikov}, A.~N. and {Bel}, J. and {Bellagamba}, F. and {Bella}, M. and {Bellini}, E. and {Benabed}, K. and {Bender}, R. and {Benevento}, G. and {Bennett}, C.~L. and {Benson}, K. and {Bergamini}, P. and {Bermejo-Climent}, J.~R. and {Bernardeau}, F. and {Bertacca}, D. and {Berthe}, M. and {Berthier}, J. and {Bethermin}, M. and {Beutler}, F. and {Bevillon}, C. and {Bhargava}, S. and {Bhatawdekar}, R. and {Bianchi}, D. and {Bisigello}, L. and {Biviano}, A. and {Blake}, R.~P. and {Blanchard}, A. and {Blazek}, J. and {Blot}, L. and {Bosco}, A. and {Bodendorf}, C. and {Boenke}, T. and {B{\"o}hringer}, H. and {Boldrini}, P. and {Bolzonella}, M. and {Bonchi}, A. and {Bonici}, M. and {Bonino}, D. and {Bonino}, L. and {Bonvin}, C. and {Bon}, W. and {Booth}, J.~T. and {Borgani}, S. and {Borlaff}, A.~S. and {Borsato}, E. and {Bose}, B. and {Botticella}, M.~T. and {Boucaud}, A. and {Bouche}, F. and {Boucher}, J.~S. and {Boutigny}, D. and {Bouvard}, T. and {Bouwens}, R. and {Bouy}, H. and {Bowler}, R.~A.~A. and {Bozza}, V. and {Bozzo}, E. and {Branchini}, E. and {Brando}, G. and {Brau-Nogue}, S. and {Brekke}, P. and {Bremer}, M.~N. and {Brescia}, M. and {Breton}, M.-A. and {Brinchmann}, J. and {Brinckmann}, T. and {Brockley-Blatt}, C. and {Brodwin}, M. and {Brouard}, L. and {Brown}, M.~L. and {Bruton}, S. and {Bucko}, J. and {Buddelmeijer}, H. and {Buenadicha}, G. and {Buitrago}, F. and {Burger}, P. and {Burigana}, C. and {Busillo}, V. and {Busonero}, D. and {Cabanac}, R. and {Cabayol-Garcia}, L. and {Cagliari}, M.~S. and {Caillat}, A. and {Caillat}, L. and {Calabrese}, M. and {Calabro}, A. and {Calderone}, G. and {Calura}, F. and {Camacho Quevedo}, B. and {Camera}, S. and {Campos}, L. and {Ca{\~n}as-Herrera}, G. and {Candini}, G.~P. and {Cantiello}, M. and {Capobianco}, V. and {Cappellaro}, E. and {Cappelluti}, N. and {Cappi}, A. and {Caputi}, K.~I. and {Cara}, C. and {Carbone}, C. and {Cardone}, V.~F. and {Carella}, E. and {Carlberg}, R.~G. and {Carle}, M. and {Carminati}, L. and {Caro}, F. and {Carrasco}, J.~M. and {Carretero}, J. and {Carrilho}, P. and {Carron Duque}, J. and {Carry}, B.},
        title = "{Euclid: I. Overview of the Euclid mission}",
      journal = {\aap},
         year = 2025,
        month = may,
       volume = {697},
          eid = {A1},
        pages = {A1},
          doi = {10.1051/0004-6361/202450810},
archivePrefix = {arXiv},
       eprint = {2405.13491},
 primaryClass = {astro-ph.CO},
       adsurl = {https://ui.adsabs.harvard.edu/abs/2025A&A...697A...1E}
}

@ARTICLE{Carlsten2021ApJ...922..267C,
       author = {{Carlsten}, Scott G. and {Greene}, Jenny E. and {Greco}, Johnny P. and {Beaton}, Rachael L. and {Kado-Fong}, Erin},
        title = "{Structures of Dwarf Satellites of Milky Way-like Galaxies: Morphology, Scaling Relations, and Intrinsic Shapes}",
      journal = {\apj},
         year = 2021,
        month = dec,
       volume = {922},
       number = {2},
          eid = {267},
        pages = {267},
          doi = {10.3847/1538-4357/ac2581},
archivePrefix = {arXiv},
       eprint = {2105.03435},
 primaryClass = {astro-ph.GA},
       adsurl = {https://ui.adsabs.harvard.edu/abs/2021ApJ...922..267C}
}

@article{McConnachie_2012,
doi = {10.1088/0004-6256/144/1/4},
url = {https://doi.org/10.1088/0004-6256/144/1/4},
year = {2012},
month = {jun},
publisher = {The American Astronomical Society},
volume = {144},
number = {1},
pages = {4},
author = {McConnachie, Alan W.},
title = {THE OBSERVED PROPERTIES OF DWARF GALAXIES IN AND AROUND THE LOCAL GROUP},
journal = {The Astronomical Journal}
}

@ARTICLE{2015ApJ...798L..45V,
       author = {{van Dokkum}, Pieter G. and {Abraham}, Roberto and {Merritt}, Allison and {Zhang}, Jielai and {Geha}, Marla and {Conroy}, Charlie},
        title = "{Forty-seven Milky Way-sized, Extremely Diffuse Galaxies in the Coma Cluster}",
      journal = {\apjl},
         year = 2015,
        month = jan,
       volume = {798},
       number = {2},
          eid = {L45},
        pages = {L45},
          doi = {10.1088/2041-8205/798/2/L45},
archivePrefix = {arXiv},
       eprint = {1410.8141},
 primaryClass = {astro-ph.GA},
       adsurl = {https://ui.adsabs.harvard.edu/abs/2015ApJ...798L..45V}
}

@ARTICLE{2021A&A...656A..44R,
       author = {{Rom{\'a}n}, Javier and {Castilla}, Aida and {Pascual-Granado}, Javier},
        title = "{Discovery and analysis of low-surface-brightness galaxies in the environment of NGC 1052}",
      journal = {\aap},
         year = 2021,
        month = dec,
       volume = {656},
          eid = {A44},
        pages = {A44},
          doi = {10.1051/0004-6361/202142161},
archivePrefix = {arXiv},
       eprint = {2110.09527},
 primaryClass = {astro-ph.GA},
       adsurl = {https://ui.adsabs.harvard.edu/abs/2021A&A...656A..44R}
}

@ARTICLE{2012ApJS..200....4F,
       author = {{Ferrarese}, Laura and {C{\^o}t{\'e}}, Patrick and {Cuillandre}, Jean-Charles and {Gwyn}, S.~D.~J. and {Peng}, Eric W. and {MacArthur}, Lauren A. and {Duc}, Pierre-Alain and {Boselli}, A. and {Mei}, Simona and {Erben}, Thomas and {McConnachie}, Alan W. and {Durrell}, Patrick R. and {Mihos}, J. Christopher and {Jord{\'a}n}, Andr{\'e}s and {Lan{\c{c}}on}, Ariane and {Puzia}, Thomas H. and {Emsellem}, Eric and {Balogh}, Michael L. and {Blakeslee}, John P. and {van Waerbeke}, Ludovic and {Gavazzi}, Rapha{\"e}l and {Vollmer}, Bernd and {Kavelaars}, J.~J. and {Woods}, David and {Ball}, Nicholas M. and {Boissier}, S. and {Courteau}, St{\'e}phane and {Ferriere}, E. and {Gavazzi}, G. and {Hildebrandt}, Hendrik and {Hudelot}, P. and {Huertas-Company}, M. and {Liu}, Chengze and {McLaughlin}, Dean and {Mellier}, Y. and {Milkeraitis}, Martha and {Schade}, David and {Balkowski}, Chantal and {Bournaud}, Fr{\'e}d{\'e}ric and {Carlberg}, R.~G. and {Chapman}, S.~C. and {Hoekstra}, Henk and {Peng}, Chien and {Sawicki}, Marcin and {Simard}, Luc and {Taylor}, James E. and {Tully}, R. Brent and {van Driel}, Wim and {Wilson}, Christine D. and {Burdullis}, Todd and {Mahoney}, Billy and {Manset}, Nadine},
        title = "{The Next Generation Virgo Cluster Survey (NGVS). I. Introduction to the Survey}",
      journal = {\apjs},
         year = 2012,
        month = may,
       volume = {200},
       number = {1},
          eid = {4},
        pages = {4},
          doi = {10.1088/0067-0049/200/1/4},
       adsurl = {https://ui.adsabs.harvard.edu/abs/2012ApJS..200....4F}
}

@ARTICLE{2024AJ....168...69Z,
       author = {{Zaritsky}, Dennis and {Golini}, Giulia and {Donnerstein}, Richard and {Trujillo}, Ignacio and {Akhlaghi}, Mohammad and {Chamba}, Nushkia and {D'Onofrio}, Mauro and {Eskandarlou}, Sepideh and {Hosseini-ShahiSavandi}, S. Zahra and {Infante-Sainz}, Ra{\'u}l and {Martin}, Garreth and {Montes}, Mireia and {Rom{\'a}n}, Javier and {Sedighi}, Nafise and {Sharbaf}, Zahra},
        title = "{LIGHTS. Survey Overview and a Search for Low Surface Brightness Satellite Galaxies}",
      journal = {\aj},
         year = 2024,
        month = aug,
       volume = {168},
       number = {2},
          eid = {69},
        pages = {69},
          doi = {10.3847/1538-3881/ad543f},
archivePrefix = {arXiv},
       eprint = {2406.01912},
 primaryClass = {astro-ph.GA},
       adsurl = {https://ui.adsabs.harvard.edu/abs/2024AJ....168...69Z}
}

@ARTICLE{2016MNRAS.456.1359F,
       author = {{Fliri}, J{\"u}rgen and {Trujillo}, Ignacio},
        title = "{The IAC Stripe 82 Legacy Project: a wide-area survey for faint surface brightness astronomy}",
      journal = {\mnras},
         year = 2016,
        month = feb,
       volume = {456},
       number = {2},
        pages = {1359-1373},
          doi = {10.1093/mnras/stv2686},
archivePrefix = {arXiv},
       eprint = {1603.04474},
 primaryClass = {astro-ph.GA},
       adsurl = {https://ui.adsabs.harvard.edu/abs/2016MNRAS.456.1359F}
}

@ARTICLE{2022NatAs...6..897S,
       author = {{Sales}, Laura V. and {Wetzel}, Andrew and {Fattahi}, Azadeh},
        title = "{Baryonic solutions and challenges for cosmological models of dwarf galaxies}",
      journal = {Nature Astronomy},
         year = 2022,
        month = jun,
       volume = {6},
        pages = {897-910},
          doi = {10.1038/s41550-022-01689-w},
archivePrefix = {arXiv},
       eprint = {2206.05295},
 primaryClass = {astro-ph.GA},
       adsurl = {https://ui.adsabs.harvard.edu/abs/2022NatAs...6..897S}
}

@ARTICLE{2020ApJ...890..128F,
       author = {{Ferrarese}, Laura and {C{\^o}t{\'e}}, Patrick and {MacArthur}, Lauren A. and {Durrell}, Patrick R. and {Gwyn}, S.~D.~J. and {Duc}, Pierre-Alain and {S{\'a}nchez-Janssen}, R{\'u}ben and {Santos}, Matthew and {Blakeslee}, John P. and {Boselli}, Alessandro and {Boyer}, Fred and {Cantiello}, Michele and {Courteau}, St{\'e}phane and {Cuillandre}, Jean-Charles and {Emsellem}, Eric and {Erben}, Thomas and {Gavazzi}, Giuseppe and {Guhathakurta}, Puragra and {Huertas-Company}, Marc and {Jord{\'a}n}, Andr{\'e}s and {Lan{\c{c}}on}, Ariane and {Liu}, Chengze and {Mei}, Simona and {Mihos}, J. Christopher and {Peng}, Eric W. and {Puzia}, Thomas H. and {Roediger}, Joel and {Schade}, David and {Taylor}, James E. and {Toloba}, Elisa and {Zhang}, Hongxin},
        title = "{The Next Generation Virgo Cluster Survey (NGVS). XIV. The Discovery of Low-mass Galaxies and a New Galaxy Catalog in the Core of the Virgo Cluster}",
      journal = {\apj},
         year = 2020,
        month = feb,
       volume = {890},
       number = {2},
          eid = {128},
        pages = {128},
          doi = {10.3847/1538-4357/ab339f},
       adsurl = {https://ui.adsabs.harvard.edu/abs/2020ApJ...890..128F}
}

@ARTICLE{2022A&A...667A..76J,
       author = {{Junais} and {Boissier}, S. and {Boselli}, A. and {Ferrarese}, L. and {C{\^o}t{\'e}}, P. and {Gwyn}, S. and {Roediger}, J. and {Lim}, S. and {Peng}, E.~W. and {Cuillandre}, J.-C. and {Longobardi}, A. and {Fossati}, M. and {Hensler}, G. and {Koda}, J. and {Bautista}, J. and {Boquien}, M. and {Ma{\l}ek}, K. and {Amram}, P. and {Roehlly}, Y.},
        title = "{A Virgo Environmental Survey Tracing Ionised Gas Emission (VESTIGE). XIII. The role of ram-pressure stripping in transforming the diffuse and ultra-diffuse galaxies in the Virgo cluster}",
      journal = {\aap},
         year = 2022,
        month = nov,
       volume = {667},
          eid = {A76},
        pages = {A76},
          doi = {10.1051/0004-6361/202244237},
archivePrefix = {arXiv},
       eprint = {2208.02634},
 primaryClass = {astro-ph.GA},
       adsurl = {https://ui.adsabs.harvard.edu/abs/2022A&A...667A..76J}
}

\begin{appendix}

\section{GALFIT and surface brightness profile}\label{section_appendix:galfit}
\begin{figure*}[!t]
    \centering
    {\includegraphics [width=0.75\linewidth]
    {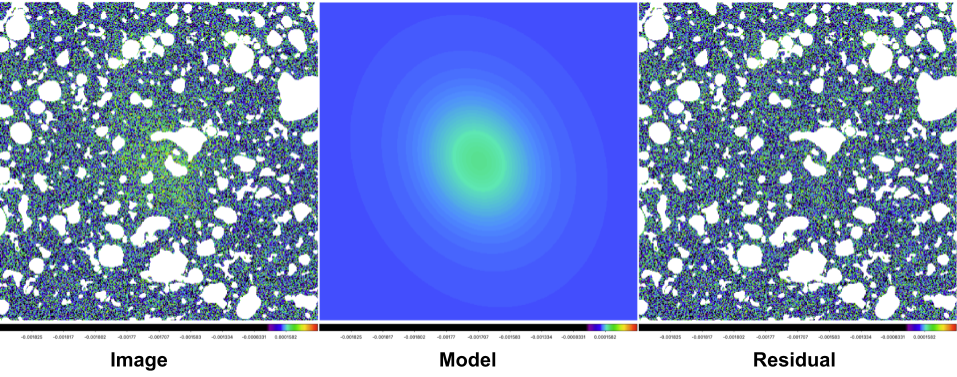}}
      \caption{Almost-dark galaxy TTT~J1237327+143535 in the $g'$ band (\textit{left}), its model in GALFIT (\textit{center}), and the residual (\textit{right}).  }
    \label{fig:galfit output dark dwarf}
    \end{figure*}

\begin{figure*}[!t]
    \centering
    {\includegraphics [width=0.75\linewidth]
    {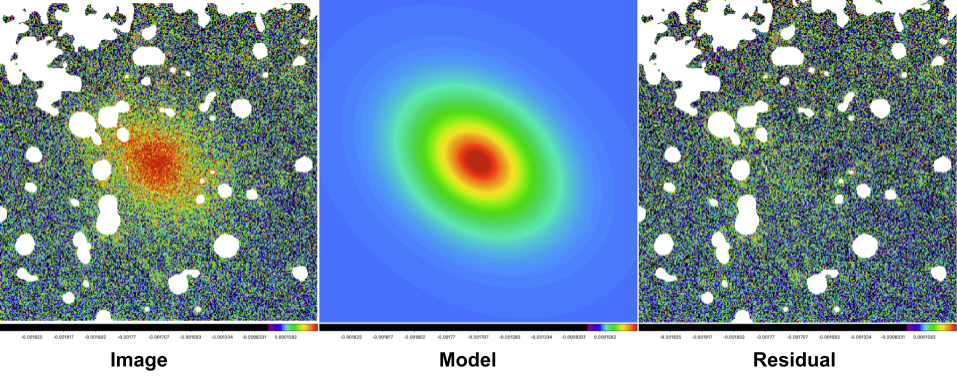}}
      \caption{UDG neighbour SMDG~J1237307+143906 in the $g'$ band (\textit{left}), its model in GALFIT (\textit{center}), and the residual (\textit{right}).}
    \label{fig:galfit output bright dwarf}
    \end{figure*}

We used GALFIT to fit the $g'$ band images and find the effective radius, position angle, ellipticity, and Sérsic index of the almost-dark galaxy TTT~J1237327+143535 (Fig~\ref{fig:galfit output dark dwarf}) and its neighbour SMDG~J1237307+143906 (Fig~\ref{fig:galfit output bright dwarf}). Because the $g'$ band is deeper than the $r'$ band, we report $g'$ band values; $r'$ band values are consistent within error bars. Properties are summarized in Table~\ref{tab:dwarfs properties}.

\begin{figure}[ht!]
    \centering
    {\includegraphics[width=1\linewidth] {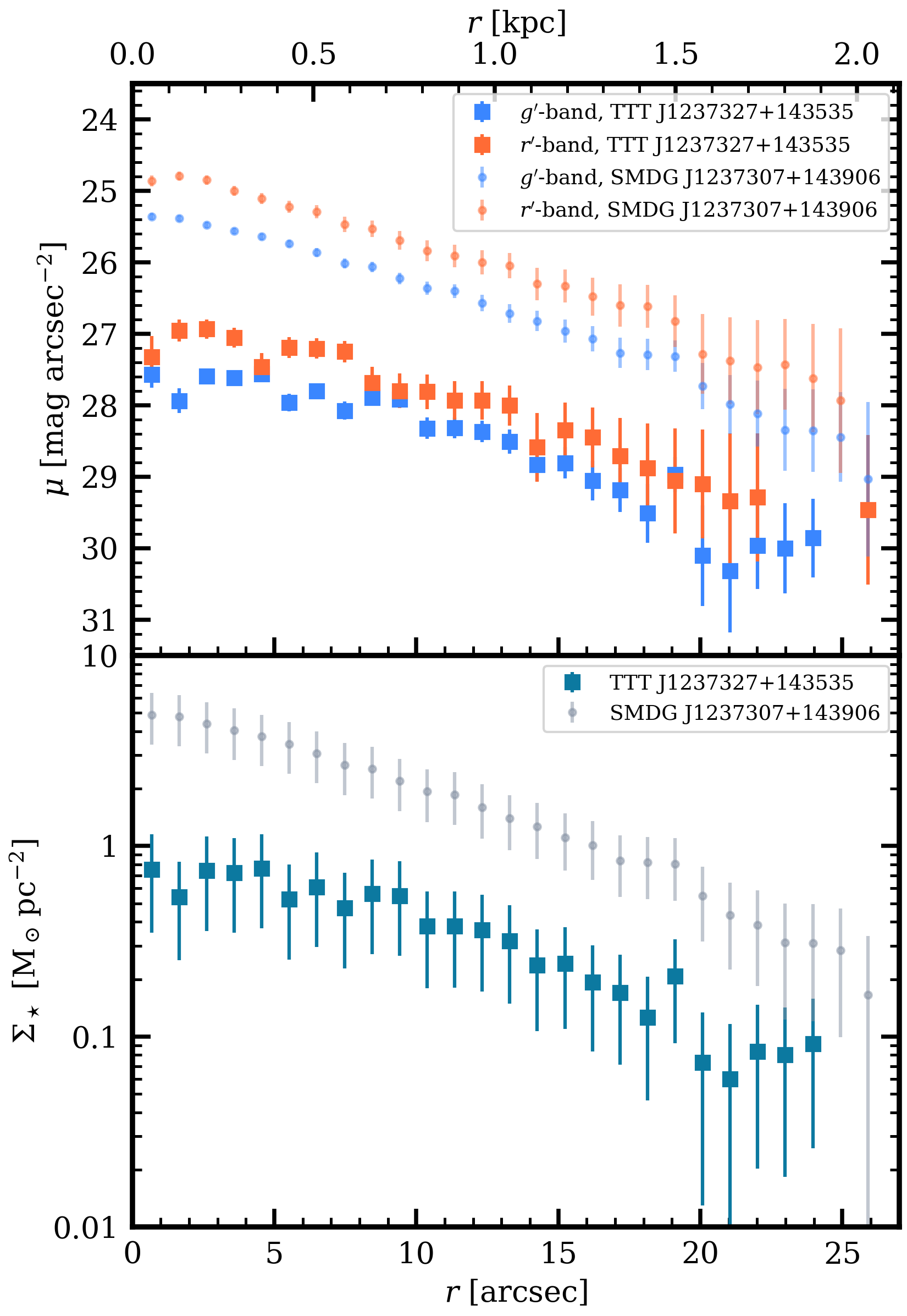}}
      \caption{\textit{Top:} Surface brightness profiles in the $g'$ band (blue) and $r'$ band (orange) of the almost-dark galaxy TTT~J1237327+143535 (squares), compared to its UDG neighbour SMDG~J1237307+143906 (circles). The surface brightness profiles are corrected for Galactic extinction. \textit{Bottom:} Surface mass density profiles of the almost-dark galaxy (squares) and its UDG neighbour (circles).}
    \label{fig:profiles}
    \end{figure}
    
\begin{table*}
\centering
\caption{ Properties\tablefoottext{a}{} of the almost-dark galaxy TTT~J1237327+143535 and its neighbour SMDG~J1237307+143906. }
\begin{tabular}{lccc}
\hline\hline \\[0.01cm]
   & TTT~J1237327+143535 & SMDG~J1237307+143906 & SMDG~J1237307+143906 \tablefoottext{b}{} \\[0.01cm] 
   \hline 
 RA [$^\circ$] & $189.3861$& $189.3779$ & $189.3779$\\ 
 DEC [$^\circ$] & $14.5932$ &  $14.6517$ & $14.6517$\\
 $r_\mathrm{e}$ [$''$] & $11.9\pm 0.5$ & $12.2 \pm 0.1$  & $12.99^{+3.25}_{-0.52}$\\
 $r_\mathrm{e}$ [$\kpc{}$] & $0.93\pm 0.04$ & $0.96 \pm 0.01$  & --\\
 $n$  &  $0.54\pm 0.03$ & $ 0.75 \pm 0.01$ & $0.85_{-0.04}^{+0.17}$\\
 $b/a$ & $0.81 \pm 0.03$ &  $0.74 \pm 0.01$ & $0.76 \pm 0.04$ \\ 
 PA [$^\circ$] &  $26.2 \pm 6.5$ &  $53.1 \pm 0.9$ & $43.5\pm 5.77$\\
 $\mu_{0,g}$ [$\magarcsec$]  & $27.88 \pm 0.15$& $25.36 \pm 0.02$ & $25.31_{-0.23}^{+0.05}$\\
 $\mu_{0,r}$ [$\magarcsec$]  & $26.93 \pm 0.11$ & $24.80 \pm 0.03$ & $24.70_{-0.23}^{+0.05}$\\
 $\left\langle\mu_{g}\right\rangle_\mathrm{e}$ [$\magarcsec$] & $28.14 \pm 0.11$ & $26.08 \pm 0.02$ & --\\
 $\left\langle\mu_{r}\right\rangle_\mathrm{e}$ [$\magarcsec$] & $27.52 \pm 0.11$ & $25.44 \pm 0.02$ & --\\
 $(g'-r')_0$ \tablefoottext{c}{} &  $0.60 \pm 0.10$ & $0.58 \pm 0.04$ & --\\
 $g$ [mag]  & $21.07 \pm 0.04$ & $19.06 \pm 0.01$ & $18.91_{-0.41}^{+0.05}$\\
 $r$ [mag] & $20.43\pm 0.04$ & $18.40 \pm 0.01$ & $18.30_{-0.36}^{+0.05}$\\
 $M_\star$ $[\mathrm{M_\odot}]$ &  $(2.16 \pm 0.38) \times 10^6$ & $(1.02 \pm 0.10) \times 10^7$ & -- \\
\hline
\end{tabular}
\label{tab:dwarfs properties}
\tablefoot{\tablefoottext{a}{The effective radius $r_\mathrm{e}$, Sérsic index $n$, axis ratio $b/a$, position angle PA, and total magnitude in both bands are from the fit results using GALFIT. The central and mean effective surface brightness values are corrected for Galactic extinction.}
\tablefoottext{b}{Results by \cite{2023ApJS..267...27Z} using data from DECaLS, show consistency with our measurements of SMDG~J1237307+143906.}
\tablefoottext{c}{Median Galactic extinction correct $(g'-r')_0$ colour within $r_\mathrm{e}$ of the galaxies and bootstrap errors.}
   
}
\end{table*}

\end{appendix}
\end{document}